\documentclass[twocolumn]{jpsj3} %% two-column layout
\usepackage{graphicx}% Include figure files
\usepackage{dcolumn}% Align table columns on decimal point
\usepackage{bm}% bold math

\renewcommand{\bm}[1]{\mbox{\boldmath$#1$}}

\title{Weak Spin Fluctuation with Finite Wave Vector and \\
Superconducting Gap Symmetry in K$_x$Fe$_{2-y}$Se$_2$: \\
$^{77}$Se Nuclear Magnetic Resonance }

\author{Hisashi \textsc{Kotegawa}$^{1,3}$\thanks{E-mail address: kotegawa@crystal.kobe-u.ac.jp}, Yusuke \textsc{Tomita}$^{1}$, Hideki \textsc{Tou}$^{1,3}$, \\ Yoshikazu \textsc{Mizuguchi}$^{2,3}$, Hiroyuki \textsc{Takeya}$^{2,3}$, and Yoshihiko \textsc{Takano}$^{2,3}$}

\inst{$^{1}$Department of Physics, Kobe University, Kobe 658-8530, Japan\\
$^{2}$National Institute for Materials Science, 1-2-1, Sengen, Tsukuba 305-0047, Japan\\
$^{3}$JST, Transformative Research-Project on Iron Pnictides (TRIP), Chiyoda, Tokyo 102-0075, Japan\\

}

\abst{
We report $^{77}$Se-nuclear magnetic resonance (NMR) results down to sufficiently low temperatures under magnetic fields parallel to both the $ab$-plane and the $c$-axis in a paramagnetic/superconducting (PM/SC) phase of K$_x$Fe$_{2-y}$Se$_2$. 
The observation of anisotropy in the orbital part of the Knight shift results in the anisotropy of its spin part increasing on approaching the transition temperature.
The anisotropy of the Korringa relation suggests the presence of the weak spin fluctuations with a finite wave vector $\bm{q}$, which induce the magnetic fluctuations along the $ab$-plane at the Se site.
Such fluctuations do not correspond to the stripe $(\pi,0)$ correlation of the Fe moment observed in many Fe-based superconductors, and are not contradictory to weak $(\pi,\pi)$ correlations.
The nuclear spin-lattice relaxation rate $1/T_1$ shows a field-independent $T_1T \sim const.$ behavior at low temperatures for $H \parallel ab$, which  is attributed to the nonzero density of states at the Fermi level and can be explained by the sign-changing order parameter even for nodeless gaps.
The temperature dependence of $1/T_1$ is reproduced well by nodeless models with two isotropic gaps or a single anisotropic gap.
The obtained gap magnitude in the isotropic two-gap model is comparable to those obtained in the angle-resolved photoemission spectroscopy experiments.
}

\kword{K$_{x}$Fe$_{2-y}$Se$_2$, superconductivity, NMR}

\begin{document}
\maketitle

\section{Introduction}

Fe-based superconductors contain various types of compound, and superconductivity is commonly induced at a two-dimensional layer including Fe.
However, the superconducting (SC) mechanism is not common even in a nesting scenario owing to several selectivities in the nesting vector in multi band systems.\cite{Kuroki}
The alkali-doped iron chalcogenide $A_x$Fe$_{2-y}$Se$_2$ ($A=$ K, Rb, Cs, and Tl) is noticed as an exceptional example of Fe-based superconductors.\cite{Guo}
One important feature of this material is its different Fermi surface from other Fe-based systems.
Angle-resolved photoemission spectroscopy (ARPES) studies have shown that hole pockets at the $\Gamma$ point almost disappear,\cite{Zhang,Qian} therefore it is suggested that the nesting vector is not a stripe $(\pi,0)$ but a checkerboard $(\pi,\pi)$.
As for the SC gap symmetry, the same $s^{\pm}$ wave as those of other Fe-based superconductors is excluded in this Fermi surface, and theoretically a nodeless $d$-wave, a nodal $d$-wave, and an $s$-wave have been proposed by different approaches in a magnetically mediated scenario.\cite{Wang,Das,Maier,Saito,Mazin,Fang,RYu,Lu}
On the other hand, the $s^{++}$-wave symmetry has been proposed in an orbital-fluctuation scenario.\cite{Saito}
Experimentally, the nodeless gap has been suggested by several experiments.\cite{Zhang,Qian,Zeng}
Another feature of this material is the presence of an antiferromagnetic (AF) phase with a high transition temperature, and the phase segregation between a paramagnetic (PM) phase and an AF phase.\cite{Bao,Ryan,Shermadini,Texier}
Many experimental results suggest that the PM phase is a minor phase compared with the major AF phase.\cite{Ryan,Shermadini,Texier}
Therefore, microscopic measurements such as NMR are effective for obtaining information on different phases separately.
Earlier NMR results on the PM/SC phase have been in accord on the point that spin fluctuations are weak in K$_x$Fe$_{2-y}$Se$_2$ because the nuclear spin-lattice relaxation rate $1/T_1$ is not enhanced with decreasing temperature \cite{Yu,Kotegawa,Torchetti}.
We have reported that weak AF fluctuations are enhanced towards low temperatures; however, the evaluation of the anisotropy of the fluctuations was insufficient.
As for the SC gap symmetry, $1/T_1$ in the SC state has been measured; however, it was difficult to give a conclusive remark owing to absence of the field dependence of $1/T_1$.\cite{Yu,Kotegawa}
In this study, we reexamined $^{77}$Se-NMR down to a sufficiently low temperature of 1.6 K under magnetic fields parallel to both the $ab$-plane and the $c$-axis, and also their field dependences.
We discuss the character of spin fluctuations and the SC gap symmetry from NMR results in K$_x$Fe$_{2-y}$Se$_2$.

\section{Experimental Procedure}

A single-crystalline sample with $T_c=32$ K was prepared as described elsewhere.\cite{Mizuguchi}
$^{77}$Se-NMR (the nuclear spin of $I=1/2$) measurement was performed using a standard spin-echo method under magnetic fields along the $ab$-plane and $c$-axis ($H=5$ T and 9 T).
Knight shift was obtained using a gyromagnetic ratio of $\gamma_n=8.13$ MHz/T, which is different from that used by other groups,\cite{Yu,Torchetti} giving an inconsistency in the absolute value of the Knight shift; however, it has no effect on the analysis using the spin part of the Knight shift discussed in this paper.
The nuclear spin-lattice relaxation rate $1/T_1$ was obtained by fitting the recovery curve to the single exponential function in the normal state.
In the SC state, we omitted a small amount of fast relaxation arising from the vortex core during the fitting.

\section{Experimental Results and Discussions}

\subsection{Magnetic anisotropy and spin fluctuations}

Figure 1 shows the $^{77}$Se-NMR spectra measured for $H \parallel ab$ and $H \parallel c$.
The NMR line width is sensitive to the random deficiency in ions;\cite{Nohara} however, that of K$_x$Fe$_{2-y}$Se$_2$ is comparable to that of stoichiometric FeSe$_{1.01}$ in spite of the larger Knight shift in K$_x$Fe$_{2-y}$Se$_2$.\cite{Imai}
Therefore, no distinct randomness of the crystal is confirmed in the PM/SC phase.
The spectrum for $H \parallel ab$ is split into two peaks at high temperatures owing to the difference in hyperfine coupling constant ($A_{ab}$), as reported previously.\cite{Torchetti} 
This suggests the existence of two inequivalent Se sites, whose $A_{ab}$'s are different by $\sim8$\%.
In the case of the $\sqrt{5} \times \sqrt{5}$ superstructure with an Fe vacancy order,\cite{Bao} the Se sites are divided into two sites, however each ratio is 1:4.
The present and previous experiments consistently show that the ratio is 1:1, independent of the sample, which does not correspond to the $\sqrt{5} \times \sqrt{5}$ superstructure.
The ratio of 1:1 is rather consistent with the $\sqrt{2} \times \sqrt{2}$ superstructure.\cite{Ricci,MWang,Cai}
If the superstructure has a $\sqrt{2} \times \sqrt{2}$ unit cell with respect to the original Se lattice without an Fe vacancy order, as shown in the figure, the Se sites are divided into two inequivalent sites of 1:1.\cite{Cai}

\begin{figure}[htb]
\centering
\includegraphics[width=0.8\linewidth]{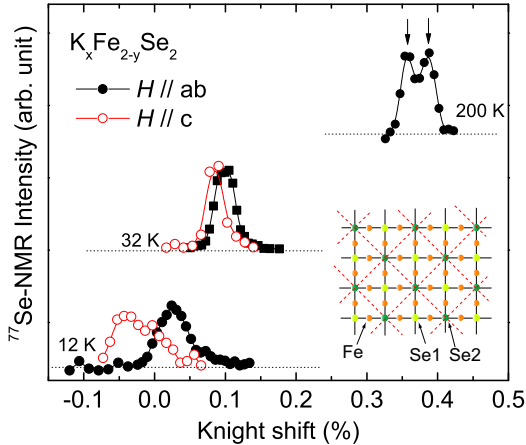}
\caption[]{(color online) $^{77}$Se-NMR spectrum attributed to the PM/SC phase for $H \parallel ab$ ($\sim8.995$ T) and $H \parallel c$ ($\sim4.995$ T). At 200 K, the signal is composed of two Se sites. At low temperatures, the spectra for $H \parallel ab$ and $H \parallel c$ are observed at different positions, indicative of the anisotropy of the orbital part of the Knight shift. The red dotted line in the crystal structure indicates the $\sqrt{2} \times \sqrt{2}$ superstructure with respect to the original Se lattice, which is a possible explanation of the two inequivalent Se sites.
}
\end{figure}

\begin{figure}[htb]
\centering
\includegraphics[width=0.8\linewidth]{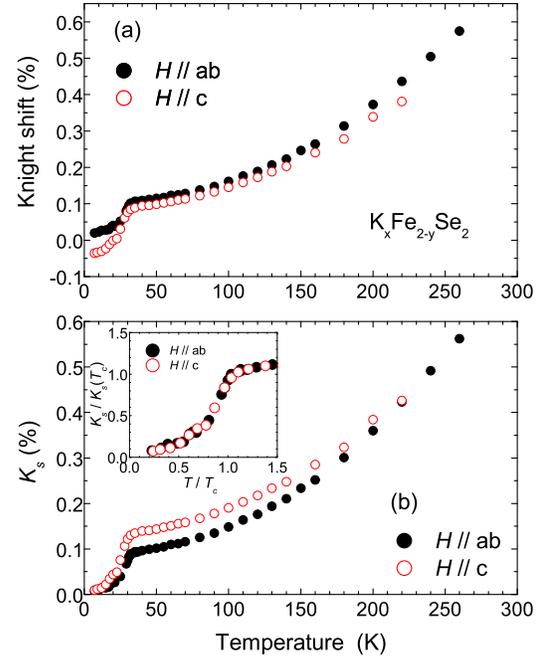}
\caption[]{(color online) Temperature dependences of (a) Knight shift and (b) spin part $K_s$ for $H \parallel ab$ and $H \parallel c$. $K_s$ becomes anisotropic with decreasing temperature. The inset shows normalized $K_s(T)$ in the SC state. 
}
\end{figure}

The spectra for $H \parallel ab$ and $H \parallel c$ are observed at almost the same position at 32 K just above $T_c$; however, they are observed at different positions at low temperatures well below $T_c$.
We estimated the Knight shift from the peak position in the SC state, because the spectrum broadens asymmetrically owing to the presence of the vortex.
At high temperatures for $H \parallel ab$, the Knight shift was determined by the center of the two peaks.
Figure 2(a) shows the temperature dependences of the Knight shift for $H \parallel ab$ and $H \parallel c$.
The Knight shift is composed of the temperature-dependent spin part $K_s(T)$ and the temperature-independent orbital (chemical) part $K_{orb}$. 
The temperature dependence of the Knight shift has already been reported,\cite{Yu,Kotegawa,Torchetti} but we newly observed the Knight shift for both fields down to low temperatures well below $T_c$.
It is apparent that the Knight shift is anisotropic well below $T_c$.
We evaluated $K_{orb}$ from the extrapolation toward $T=0$ while taking into account of a residual density of state of $6$\%, which is estimated from $1/T_1$, as mentioned below.
Each orbital part is estimated to be $K_{orb}^{ab}=0.018$\% and $K_{orb}^{c}=-0.045$\%.
The temperature dependences of $K_s$, which is obtained by subtracting $K_{orb}$ from the total Knight shift, are displayed in Fig.~2(b).
$K_s^c$ at $T_c$ is estimated to be $\sim0.14$\%, which is in good agreement with the previous report measured at 8.3 T.\cite{Torchetti}
$K_s$ is isotropic at high temperatures; however, the anisotropy is induced with decreasing temperature.
Since $K_s^{i} = A_{i} \chi^{i}(\bm{q}=0,\omega=0)$ ($i=$ $ab$ or $c$), the spin susceptibility $\chi(0,0)$ along the $ab$-plane is strongly suppressed with decreasing temperature.
The anisotropy $K_s^{c}/K_s^{ab} \sim 1.45$ at $T_c$ is opposite to the As sites in BaFe$_2$As$_2$, SrFe$_2$As$_2$, and LaFeAs(O,F), where both $K_s^{c}/K_s^{ab}$ and $\chi^{c}/\chi^{ab}$ are less than 1.\cite{Kitagawa1,Kitagawa2,Kitagawa3}
$K_s^{c}/K_s^{ab} > 1$ is observed at the Te site in Fe(Te,Se).\cite{Michioka}
These differences are conjectured to be induced by the difference between the magnetic properties and/or the hybridizations of Fe-As and Fe-Se(Te).

\begin{figure}[htb]
\centering
\includegraphics[width=0.8\linewidth]{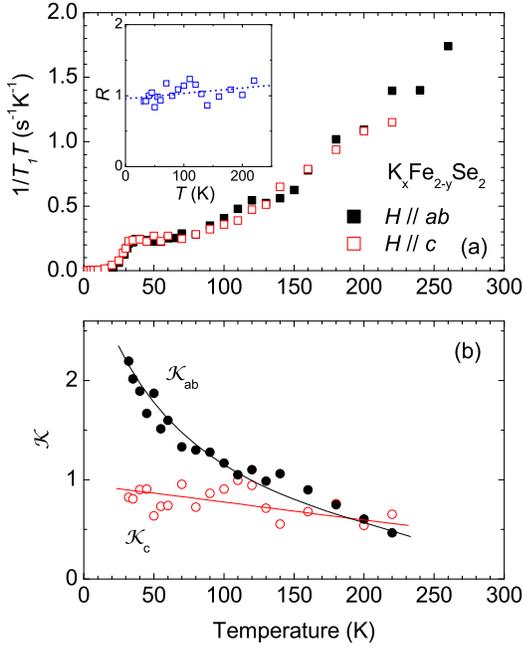}
\caption[]{(color online) (a) Temperature dependences of $1/T_1T$ for $H \parallel ab$ and $H \parallel c$. The inset shows $R \equiv (1/T_1T)_{H \parallel ab} / (1/T_1T)_{H \parallel c}$. The dotted line is obtained by the least-squares method. (b) Temperature dependences of Korringa ratio $\texttt{K}$ for each direction. The anisotropy of $\texttt{K}$ is enhanced with decreasing temperature.
}
\end{figure}

Figure 3(a) shows the temperature dependences of $1/T_1T$ for $H \parallel ab$ and $H \parallel c$.
We cannot observe anisotropy in $1/T_1T$, consistent with a previous report.\cite{Yu}
General relations for $1/T_1T$ and $K_s$ are given as $1/T_1T \propto \sum_q |A_q|^2 \chi"_{\perp}(\bm{q},\omega_n)$ and $K_s^i = A_i \chi^i(0,0)$, where $A_q$ is a $\bm{q}$-dependent hyperfine coupling constant.
From these relations, the inconsistency between $R \equiv (1/T_1T)_{H \parallel ab} / (1/T_1T)_{H \parallel c} \sim 1$ and $K_s^{c}/K_s^{ab}$ = 1.45 implies that $\chi"(\bm{q}\neq0,\omega \sim 0)$ is anisotropic at low temperatures.
%\begin{eqnarray}
%\left( \frac{1}{T_1} \right)_{H \parallel z} = \frac{\gamma_n^2}{2}  \int_{-\in%fty}^{\infty} \{ \langle \delta H_x(t) \delta H_x(0) \rangle + \\
% \langle \delta H_y(t) \delta H_y(0) \rangle \} \exp(-i \omega_n t) dt
%\end{eqnarray}
%\begin{equation}
%\left( \frac{1}{T_1T} \right)_{H \parallel z} = \frac{\gamma_n^2 k_B}{(\gamma_e% \hbar)^2} \sum_{\bm{q}} \left( |A_q^x|^2 \frac{\chi''_x(\bm{q},\omega_n)}{\ome%ga_n} + |A_q^y|^2 \frac{\chi''_y(\bm{q},\omega_n)}{\omega_n} \right)
%\end{equation}
%where $\langle \delta H_x(t) \delta H_x(0) \rangle$ corresponds to the magnetic% fluctuations along $x$-axis induced at the Se site.
We obtained the Korringa ratio taking anisotropy into account as follows:
\begin{equation}
\texttt{K}^{\ i}=\left( \frac{1}{T_1T} \right)^{i} \left( \frac{1}{K_s^{i}} \right)^2 \frac{\hbar}{4\pi k_B} \frac{\gamma_e^2}{\gamma_n^2} \ \ \  (i =  ab \ {\rm or} \ c),
\end{equation}
where $(1/T_1T)^{ab} =(1/T_1T)_{H \parallel c}$ and $(1/T_1T)^{c} = 2(1/T_1T)_{H \parallel ab} - (1/T_1T)_{H \parallel c}$ obtained from $1/T_1T \propto \sum_q |A_q|^2 \chi"_{\perp}(\bm{q},\omega_n)$.
If $\texttt{K}$ increases with decreasing temperature, this means that spin fluctuations at $\bm{q} \neq 0$ are developing.
Figure 3(b) shows the temperature dependence of the Korringa ratio $\texttt{K}^{\ i}$ $(i=ab$ and $c)$.
The temperature dependence of $\texttt{K}^{\ c}$ is weak; however, $\texttt{K}^{\ ab}$ increases toward $T_c$, thereby suggesting that $\bm{q} \neq 0$ spin fluctuations survive at low temperatures in spite of the strong suppression of the spin susceptibility $\chi(0,0)$.
$\texttt{K}^{\ ab}$ is almost the same as that in our previous report,\cite{Kotegawa} but Korringa relations reported by other groups rather correspond to $\texttt{K}^{\ c}$.\cite{Yu,Torchetti}
The anisotropy of $\texttt{K}^{\ i}$ suggests that the magnetic fluctuations along the $ab$-plane are induced at the Se site.
%This is contrast to the stripe $(\pi,0)$ fluctuation where the magnetic fluctua%tions along the $c$-axis is stronger than that along the $ab$ plane at the As s%ite.\cite{Kitagawa1}

The temperature dependence of $R \equiv (1/T_1T)_{H \parallel ab} / (1/T_1T)_{H \parallel c}$ is shown in the inset of Fig.~3(a).
In some Fe-based superconductors, $R \sim 1.5$ is observed to originate from the stripe $(\pi,0)$ fluctuation.\cite{Kitagawa1,Kitagawa2,Kitagawa3,Li}
In K$_x$Fe$_{2-y}$Se$_2$, $R$ is almost 1 in a wide temperature range.
The internal field at the Se site induced by four neighboring Fe moments connects to each Fe-spin component $S_i$ via the hyperfine coupling tensor as follows, while neglecting the in-plane anisotropy:\cite{Kitagawa1} \\

\noindent
For $\bm{Q}=0$,
\begin{eqnarray}
\bm{H}_{hf}^{Se} = \tilde{A} \bm{S}
=\left(
\begin{array}{ccc}
A_{ab} & 0 & 0 \\
0 & A_{ab} & 0 \\
0 & 0 & A_c \\
\end{array}
\right)
\left(
\begin{array}{c}
S_{ab} \\
S_{ab} \\
S_c \\
\end{array}
\right).
\end{eqnarray}

\noindent
For, $\bm{Q}=(\pi,0)$ or $(0,\pi)$ \\
\begin{eqnarray}
\bm{H}_{hf}^{Se} = \tilde{A} \bm{S}
=\left(
\begin{array}{ccc}
0 & 0 & B_1 \\
0 & 0 & B_2 \\
B_1 & B_2 & 0 \\
\end{array}
\right)
\left(
\begin{array}{c}
S_{ab} \\
S_{ab} \\
S_c \\
\end{array}
\right). 
\end{eqnarray}

\noindent
For $\bm{Q}=(\pi,\pi)$,
\begin{eqnarray}
\bm{H}_{hf}^{Se} = \tilde{A} \bm{S}
=\left(
\begin{array}{ccc}
0 & C & 0 \\
C & 0 & 0 \\
0 & 0 & 0 \\
\end{array}
\right)
\left(
\begin{array}{c}
S_{ab} \\
S_{ab} \\
S_c \\
\end{array}
\right).
\end{eqnarray}  
Here, the diagonal parts correspond to the local $\bm{Q}=0$ arrangement, and $B_{1,2}$ [$C$] corresponds to the stripe $\bm{Q}=(\pi,0)$ or $(0,\pi)$ [checkerboard $\bm{Q}=(\pi,\pi)$] arrangements.
If the Fe-spin component is fluctuating with a specific correlation as $S_{i}(\bm{q},\omega)$, it induces magnetic fluctuations at the Se site via this hyperfine coupling tensor.
For instance, in the $(\pi,0)$ correlation, $S_{ab}(\bm{Q},\omega)$ induces a magnetic fluctuation along the $c$-axis at the Se site, and $S_{c}(\bm{Q},\omega)$ induces the fluctuation along the $ab$ plane.
In most Fe-based superconductors possessing the $(\pi,0)$ correlation, $S_{ab}(\bm{Q},\omega_n) \geq S_{c}(\bm{Q},\omega)$ has been confirmed, which induces a stronger fluctuation along the $c$-axis at the Se site, however, $\texttt{K}^{\ ab} > \texttt{K}^{\ c}$ in K$_x$Fe$_{2-y}$Se$_2$ does not match this situation. In the checkerboard $(\pi,\pi)$ correlation, it can induce a magnetic fluctuation along the $ab$-plane at the Se site, as seen in the hyperfine coupling tensor, while the fluctuation along the $c$-axis vanishes owing to the cancellation of the contribution from four neighboring Fe moments.
This is consistent with the weak temperature dependence of $\texttt{K}^{\ c}$, and a rather remarkable development of $\texttt{K}^{\ ab}$.
We can obtain the anisotropy of $1/T_1$ from the contributions of the neighboring four Fe sites.\cite{Kitagawa2,Kitagawa3}
\begin{eqnarray}
R_{q=0} \  &=& 0.5+ 0.5\left( \frac{A_c S_{c}(0,\omega_{n})}{A_{ab}S_{ab}(0,\omega_{n})} \right)^2 \\
&\sim& 0.5+ 0.5\left( \frac{K_s^c}{K_s^{ab}} \right)^2 = 1.55, \\
R_{stripe} &=& 0.5+ \left( \frac{S_{ab}(\bm{Q},\omega_{n})}{S_c(\bm{Q},\omega_{n})} \right)^2 \sim 1.5, \\
R_{check}  &=& 0.5.
\end{eqnarray} 
The first case originates in the diagonal part of the hyperfine tensor, which is treated as an uncorrelated one if $\bm{q}$ is not identified,\cite{Kitagawa3} but here we restrict the $\bm{q}=0$ component.
If we use $A_{i}S_{i}(0,\omega_{n}) \sim K_s^i$ and $K_s^c/K_s^{ab}=1.45$, we obtain $R_{q=0} \sim 1.55$.
In the case of the $\bm{Q}=(\pi,0)$ correlation, $R_{stripe}$ becomes 1.5 by assuming $S_{ab}(\omega_{n})=S_c(\omega_{n})$.
In the case of the $\bm{Q}=(\pi,\pi)$ correlation, $R_{check}$ becomes 0.5.
The above values should be observed in typical cases where correlations are strongly developing at a specific \bm{Q}.
$R \sim 1$ in K$_x$Fe$_{1-y}$Se$_2$ indicates that spin fluctuations, at least in the low-frequency part, are not occupied strongly by the above-mentioned wave vector.
However, if correlations are moderate, the observed values would be intermediate ones.
For instance, $R_{q=0} \sim 1.55$ requires $R_{q \neq 0}<1$ for realizing the observed $R\sim1$ for $\bm{q}$-summation, which gives no contradiction to the presence of weak $(\pi,\pi)$ correlations.
Another possible explanation of $R \sim 1$ may be the $(\pi,\pi/2)$ correlation reported on the basis of an inelastic neutron scattering.\cite{Park}
In this correlation, the local arrangement of four Fe moments is a repetition of $(\pi,0)$ and $(\pi,\pi)$, which suppresses the anisotropy of $1/T_1$ and reproduces $R\sim1$.

\subsection{Superconducting gap symmetry}

\begin{figure}[htb]
\centering
\includegraphics[width=0.8\linewidth]{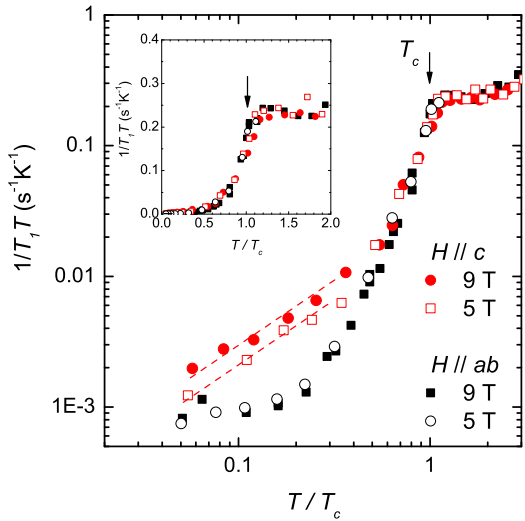}
\caption[]{(color online) Temperature dependences of $1/T_1T$ to show field-direction dependence. The small additional contribution was observed at low temperatures for $H \parallel c$. The inset shows the same data in the linear plot. The dotted lines indicate the $1/T_1 \propto T^2$ dependence.}
\end{figure}

Next, we move on to the relaxation in the SC state.
$1/T_1$ well below $T_c$ has been studied using a powdered sample \cite{Yu} and a single crystal for $H \parallel ab$ in our measurement.\cite{Kotegawa}
Our previous measurement has not been satisfactorily performed down to low temperatures; however, in this study, we measured $T_1$ down to $1.6$ K in both field directions.
Figure 4 shows the temperature dependences of $1/T_1T$.
$1/T_1$ for $H \parallel c$ has an additional contribution being distinct at low temperatures, although this is not obvious in the linear plot [inset of Fig.~4] or Knight shift [inset of Fig~2(b)].
This additional contribution roughly obeys $1/T_1 \propto T^2$, which is similar to the observation in the powdered sample.\cite{Yu}
Such anisotropic behavior against $H \parallel ab$ and $H \parallel c$ in $1/T_1$ below $T_c$ has not been reported in other Fe-based superconductors such as (Ba,K)Fe$_2$As$_2$.\cite{Li} 
A small field dependence implies that this contribution originates from the effect of the vortex core or from the modification of the SC gap by a magnetic field along the $c$-axis.
On the other hand, we should carefully consider an extrinsic contribution from the phase segregated AF phase via the spin diffusion effect; however, it would be excluded because the observed $T_1 \sim const.$ behavior in the AF phase\cite{Texier} is not seen in the present case.
The reason for $1/T_1 \propto T^2$ is not yet clear at present; however, $1/T_1T$ for $H \parallel ab$ shows no field dependence even down to the lowest temperature. 
Thus, it is suggested that $1/T_1T$ for $H \parallel ab$ is dominated by SC quasiparticles, and we applied some models on the SC symmetry to the temperature dependence of $1/T_1T$ for $H \parallel ab$.

\begin{figure}[b]
\centering
\includegraphics[width=0.8\linewidth]{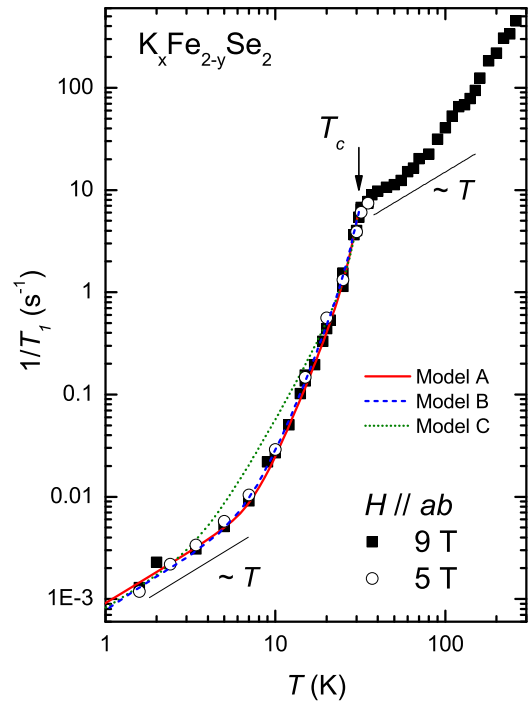}
\caption[]{(color online) Temperature dependences of $1/T_1$ to show field dependence. A field-insensitive $T_1T\sim const.$ behavior is observed below $\sim5$ K for $H \parallel ab$, confirming the nonzero DOS at the Fermi level. The red curve shows the result of a simulation using the nodeless isotropic two gaps, and the blue curve is obtained using a nodeless anisotropic single gap. The green curve indicates the nodal-gap model. Each parameter is given in Fig.~6.
}
\end{figure}

Figure 5 shows the temperature dependences of $1/T_1$ down to 1.6 K for $H \parallel ab$ and the results calculated using several models.
The field-independent $T_1T\sim const.$ behavior below $\sim5$ K indicates that  it originates in the nonzero density of states at the Fermi level.
The estimated density of states is $\sim6$\% for that just above $T_c$.
Generally the presence of a nonzero density of states is understood for impurity scattering for a nodal gap.
However, it also can be accounted for by sign-changing nodeless gaps in the presence of impurity scattering.\cite{Preosti,Parker,Bang}

\begin{figure}[htb]
\centering
\includegraphics[width=0.95\linewidth]{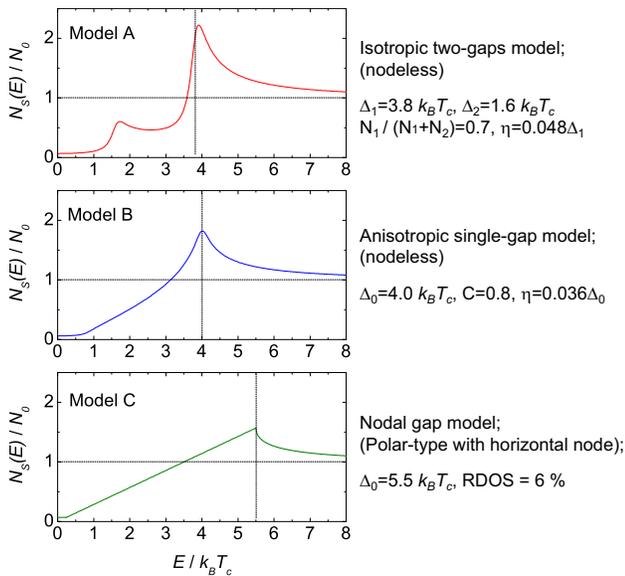}
\caption[]{(color online) Energy dependence of density of states for SC quasiparticles for three models. Each parameter was obtained to reproduce the temperature dependence of $1/T_1$.}
\end{figure}

Some theoretical works addressing the SC gap symmetry have been performed based on the band structure without a Fermi surface at the $\Gamma$ point, and a nodeless $d$-wave, a nodal $d$-wave, and an $s$-wave have been proposed in a magnetically mediated scenario.\cite{Wang,Das,Maier,Saito,Mazin,Fang,RYu,Lu}
In some nodeless models, the in-plane oscillation of the gap-magnitude is proposed.
Here we tried to use three models to reproduce the temperature dependence of $1/T_1$ in the SC state.
The curves in Fig.~5 are calculated simulations in "Models A--C".
The normalized density of states near the Fermi level for each model ($N(E)/N_0$) is shown in Fig.~6.
The red curve (Model A) indicates the isotropic two-gap model without the coherence effect.
It basically corresponds to the $s^{\pm}$-wave model,\cite{Nagai} but here the coherence effect is completely excluded.
If the coherence effect is not excluded completely by the interband mismatch, a slightly large gaps are estimated.\cite{Kotegawa}
$N_1$ and $N_2$ correspond to the density of states for each band, and $\eta$ is a smearing factor due to impurity scattering.
This model can reproduce the data well using the parameters shown in Fig.~6.
The presence of $\eta$ reproduces the nonzero density of states at the Fermi level, which is associated with $T_1T$ at low temperatures.
The nonzero density of states at the Fermi level can be explained for the sign-changing order parameter,\cite{Preosti,Parker,Bang} and the present $\eta$ is comparable to those of some other Fe-based superconductors with the isotropic multi gap.\cite{Yashima,Li_LiFeAs}
The blue curve (Model B) indicates the anisotropic single-gap model without a coherence effect.
The magnitude of the gap is given as $\Delta(\phi) = \Delta_0 (C \cos(2 \phi) + (1-C))$, and $\eta$ is also introduced as a smearing factor.
This also reproduces the data well; however, a large anisotropy of $C=0.8$ is required.
As a nodal gap (Model C), we tentatively used a Polar model with a horizontal line node corresponding to the theoretical suggestion,\cite{Saito,Mazin} and put the energy-independent residual density of states near the Fermi level, as shown in Fig.~6.
Sufficient reproduction is difficult in this model, as shown by the green curve in the figure.
In a typical line-node model, $N(E)$ is proportional to $E$ near the Fermi level, which gives $T^3$ dependence of $1/T_1$.
In the presence of a residual density of states, $1/T_1$ shows a gradual connection of the $T^3$ dependence to the $T$ dependence with decreasing temperature; however, this does not match the experimental data.
Such a temperature dependence is common in a line-node gap independent of the model, indicating that a single nodal gap with a line node should be excluded.
The two nodal gaps also did not agree with the data.
For Models A and B, the identification of the SC gap symmetry is not easy from the $1/T_1$ data; however, the results obtained using Model A are almost consistent with the ARPES results,\cite{Zhang,Xu,XPWang} where two gaps with $\Delta_1\sim10$ meV $(=3.6k_BT_c)$ and $\Delta_2\sim7$ meV $(=2.5k_BT_c)$ are suggested to open at different Fermi pockets: the $M$ and $Z$ points.
Xu {\it et al.} and Wang {\it et al.} have observed the isotropic gap even in the Fermi pocket at the $Z$ point, discarding the $d$-wave symmetry, because the node is expected to be present there in the $d$-wave symmetry.\cite{Xu,XPWang}
Unfortunately, it was difficult from $1/T_1$ to distinguish whether or not the node is present at a smaller gap in the presence of a larger isotropic gap; however, the NMR result is compatible with the multiple isotropic gaps suggested by ARPES.

\section{Conclusions}

In summary, NMR measurements in K$_x$Fe$_{2-y}$Se$_2$ were performed down to sufficiently low temperatures for both $H \parallel ab$ and $H \parallel c$, and the magnetic characteristics and SC property in the PM phase were investigated by an almost bulk-sensitive and phase-selective method.
We confirmed that superconductivity occurs in the PM phase whose crystal structure has two inequivalent Se sites, consistent with another report.
This may lead to require the refinement of some theories.
The anisotropic Korringa relation suggests the presence of a $\bm{q}\neq0$ spin fluctuation that induces a magnetic fluctuation along the $ab$-plane at the Se site.
This anisotropy cannot be explained by the stripe ($\pi,0$) correlation, and is not contradictory to weak ($\pi,\pi$) correlation.
The absence of a stripe ($\pi,0$) correlation supports the Fermi surface without pockets at the $\Gamma$ point suggested by ARPES experiments.
The temperature dependence of $1/T_1$ is reproduced well by the isotropic two-gap model with gap sizes similar to those suggested by ARPES experiments.
From an NMR point of view, the absence of a coherence effect and the presence of a nonzero density of states at the Fermi level observed in $1/T_1$ suggest the sign-changing order parameter in K$_x$Fe$_{2-y}$Se$_2$.

\section*{Acknowledgement}

The authors thank Kenji Ishida for helpful discussions.
This work has been partially supported by Grants-in-Aid for Scientific Research (Nos. 22740231, 22013011, 22340102, 24340085 and 20102005) from the Ministry of Education, Culture, Sports, Science and Technology (MEXT) of Japan.

\end{document}